\documentclass{iopart}
\usepackage{iopams}  
\usepackage{epsfig}
\begin{document}

\title[Efficient electron surfing acceleration]{Phase speed of electrostatic waves: The 
critical parameter for efficient electron surfing acceleration}


\author{M E Dieckmann$^{1}${\footnote[1]{On leave from the Department of Science and Technology, 
Linkoping University, SE-601 74 Norrkoping, Sweden}}, N J Sircombe$^{1}${\footnote[2]{On leave from the Physics Department,
University of Warwick, Coventry CV4 7AL}}, M Parviainen$^{1}$, \\
P K Shukla$^{1}$ and R O Dendy$^{3,2}$}

\address{$^{1}$ Institut f\"ur Theoretische Physik IV, Fakult\"at f\"ur
Physik und Astronomie, Ruhr-Universit\"at Bochum, 44780 Bochum, Germany}
\address{$^{2}$ Physics Department, University of Warwick, 
Coventry CV4 7AL, UK}
\address{$^{3}$ UKAEA Culham Division, Culham Science Centre, Abingdon, Oxfordshire, OX14 3DB, UK}
\eads{mardi@itn.liu.se}

\begin{abstract}
Particle acceleration by means of non-linear plasma wave interactions is of great topical interest. {Accordingly}, in this paper we focus on the electron surfing process. Self-consistent kinetic simulations, using both relativistic Vlasov and PIC (Particle In Cell) approaches, show here that electrons can be accelerated to highly relativistic energies (up to $100m_ec^2$) if the phase speed of the electrostatic wave is mildly relativistic ($0.6c$ to $0.9c$ for the magnetic field strengths considered). The acceleration is strong because of relativistic stabilisation of the nonlinearly saturated electrostatic wave, seen in both relativistic Vlasov and PIC simulations. 
An inverse power law momentum distribution can arise for the most strongly accelerated electrons. These results are of relevance to observed rapid changes in the radio synchrotron emission intensities from microquasars, gamma ray bursts and other astrophysical objects that require rapid acceleration mechanisms for electrons.
\end{abstract}



\maketitle

\section{Introduction}
Particle acceleration by means of nonlinear plasma wave interactions 
can arise from a wide variety of processes. These include wakefield 
accelerators; electron interactions with lower-hybrid wave packets 
and solitons; electron heating by collapsing electrostatic waves; and 
gyroresonant surfing acceleration  
\cite{Bingham_04a,Bingham_04b,Sircombe_05,Bingham_03,Dieckmann_00,Kuramitsu_05}. 
Here we focus on electron surfing acceleration (ESA). The principle of 
ESA is the trapping of electrons by a strong {monochromatic} electrostatic 
wave combined with the transport of the trapped electrons across 
an ambient magnetic field that is oriented orthogonally to the 
wavevector. As long as the electrons remain trapped in the potential of
the wave, they are continously accelerated, orthogonally to the
magnetic field direction and to the wave propagation direction
\cite{Sagdeev_73,Katsouleas_83}. {The resulting phase space structures are 
thus not equilibrium structures, as in an unmagnetized plasma} \cite{Luque_05}.
The electrons can be accelerated from the thermal population to energies at 
which they emit radio synchrotron radiation, in considerably less than an 
inverse ion gyroperiod \cite{Dieckmann_04a}. The rapid electron 
acceleration makes ESA a candidate mechanism for particle acceleration 
at microquasars, for example GRS 1915+105 \cite{Fender_04} and gamma 
ray bursts \cite{Piran_04}.

The electrostatic wave that traps the electrons can be excited by ion beams 
that arise ahead of shocks 
\cite{Buneman_58,Thode_73,Bauer92,Galeev_84,McClements_97}. 
As the shock expands into the upstream plasma, it reflects a substantial 
fraction of the upstream ions. If the shock normal is oriented 
perpendicular to the upstream magnetic field, ions can be specularly 
reflected \cite{Galeev_84, McClements_97} or they can undergo acceleration 
by shock surfing \cite{Shapiro_03,Ucer_01,Lee_96} and other plasma processes 
\cite{Lee_04,Lee_05}. The interplay between the plasma shock, the 
shock-reflected beam of ions and the electrons undergoing ESA has 
previously been analysed by means of kinetic particle-in-cell (PIC) 
simulations \cite{Schmitz_02a, Schmitz_02b, Shimada_00, Shimada_05, Hoshino_02}.

MHD shocks in the accretion disks of microquasars and in the outflow of 
gamma ray bursts may expand at a significant fraction of the speed of 
light $c$ into the upstream region, as MHD simulations show for
microquasars \cite{Aoki_04} and as experimental evidence indicates
for gamma ray bursts \cite{Piran_04}. The shock-reflected ions will 
move with, at least, that speed. Electrostatic waves driven by such 
ion beams have mildly relativistic phase speeds, which can increase their 
nonlinear stability and their lifetime compared to electrostatic waves 
with {large but} nonrelativistic phase speeds 
\cite{Dieckmann_00, Dieckmann_04b, Dieckmann_04c,Schamel_96}.

In this paper, we examine the efficiency of {ESA}
for waves driven by moderately dense proton beams that move at the mildly 
relativistic speeds $v_b = 0.6c$ and $v_b = 0.9c$. The efficiency with which 
the electrons are accelerated by their cross-field transport will depend critically on the wave stability; the maximum electron energy 
will be between $\approx 10$ keV found for a proton beam speed of 
$v_b \approx 0.06c$ \cite{McClements_01} and the GeV energies found for 
$v_b = 0.99c$ \cite{Dieckmann_04a}. We show below that electrons are 
accelerated up to Lorentz factors of the order of ten by ESA for $v_b=0.6c$.
The peak acceleration is further increased due to a strongly nonlinear
density modulation of one proton beam that is associated with a jump
in the beam velocity; this may evolve by a similar process to that 
reported in {Ref.} \cite{Gohda_04} for a Q-machine experiment and for the
expansion of plasma into a vacuum \cite{Sack_1987}. 
This charge accumulation yields strong localized electric fields that 
can accelerate 
electrons up to $\gamma \approx 20$, where the Lorentz factor is defined as 
$\gamma (v) = {(1-v^2/c^2)}^{-0.5}$, with $v$ {being} the electron velocity. 
The waves driven by the beam with $v_b = 0.9c$ can accelerate electrons 
to $\gamma \approx 100$, which is well above the Lorentz factors of the 
fast transient jets of the microquasar GRS 1915+105 \cite{Fender_04}, and could also account for strong synchrotron emissions. In this regime 
of beam speeds, the peak energy which the electrons can reach by ESA increases 
more than the beam speed.

The calculation of wave stability, and thus the efficiency by which electrons 
can be accelerated, should be invariant with respect to the simulation 
scheme. However, the schemes that represent the plasma vary between 
different codes and this can sometimes lead to differences in the computed 
wave stability, {e.g. due to different levels of simulation noise and 
perturbations of the islands of trapped particles} \cite{Dieckmann_04c,Schamel_96}.
The resulting different lifetimes may, in turn, affect the computed ESA efficiency. 
We, therefore, compare in section 2 the development of the beam instability in a 
PIC simulation \cite{Eastwood_91} with that in an electrostatic Vlasov simulation 
\cite{Sircombe_05,Arber_02} for an unmagnetized plasma {composed} of ions and electrons. 
The PIC code advances a number of macro-particles along the characteristics of motion, given by
		\begin{equation}
			\label{pic1}
				\frac{\partial x}{\partial t} = \frac{p}{\gamma m_e},
		\end{equation}
		\begin{equation}
			\label{pic2}
				\frac{\partial p}{\partial t} = -e E,
		\end{equation}
		for the electrons, and by
		\begin{equation}
			\label{pic3}
				\frac{\partial x}{\partial t} = \frac{p}{\gamma m_i},
		\end{equation}
		\begin{equation}
			\label{pic4}
				\frac{\partial p}{\partial t} = e E,
		\end{equation}
		for the protons.
The Vlasov code solves the one dimensional relativistic Vlasov-Poisson system directly. This fully nonlinear self consistent system is governed by the Vlasov equation for the electron distribution function $f_e (x,p,t)$ 
	\begin{equation}
			\label{vlasov_e}
			\frac {\partial f_e}{ \partial t} + \frac{p}{m_e\gamma} \frac{\partial f_e}{\partial x} - eE\frac{\partial f_e}{\partial p} = 0,
		\end{equation}
	the Vlasov equation for the ion distribution function $f_i (x,p,t)$
		\begin{equation}
			\label{vlasov_i}
			\frac {\partial f_i}{ \partial t} + \frac{p}{m_i\gamma} \frac{\partial f_i}{\partial x} + eE\frac{\partial f_i}{\partial p} = 0,
		\end{equation}
	and the Poisson equation for the electric field
		\begin{equation}
			\label{poisson}
			\frac{\partial E}{\partial x} = -\frac{e}{\epsilon_0} \left( \int f_e dv - \int f_i dv \right).
		\end{equation}
Comparison of the two codes confirms almost identical nonlinear evolution of the electrostatic waves. We then perform electromagnetic PIC simulations with different strengths of the magnetic field in section  3, and discuss the results
{in section 4}.

\section{Simulation setup and code comparison} 
\begin{figure}
\begin{center}
\epsfxsize = 8.8cm
\epsfbox{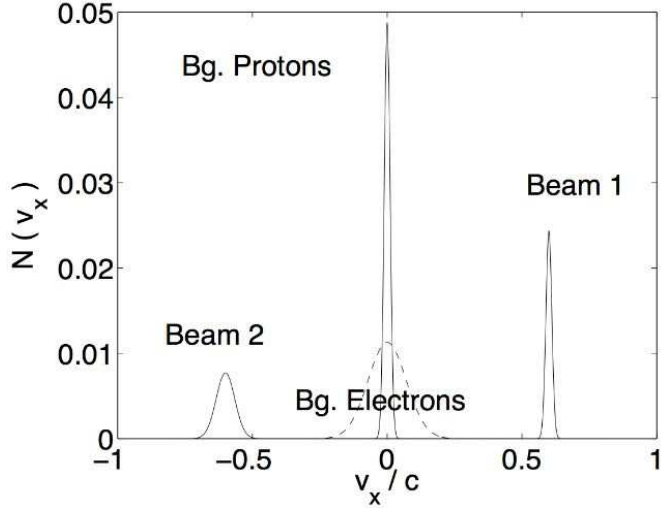}
\end{center}
\caption{\label{Fig1}
Schematic of the initial plasma distribution in our simulation.
We model two beams that have an equal mean speed and move in 
opposite directions. The thermal spread of the background 
electrons and protons and that of both beams is small 
compared to the beam speed modulus. 
Beam 2 is hotter to account for the proton beam interaction with 
the upstream medium (the thermal spread in the simulations
is less than shown here.). The initial mean speeds along $v_y$ and $v_z$
are zero for all particle species. The total proton density is equal 
to the electron density.}
\end{figure}

As a perpendicular shock propagates into the upstream medium, beams 
are formed by the shock reflected protons. Ideally one would simulate the formation of these beams, but to approach this complete problem is, at the present time, too computationally demanding. Thus, we focus instead on a one-dimensional periodic system, in which the counter-propagating beams are present at $t=0$, the evolution of which is governed by the Buneman-type instabilities \cite{Buneman_58,Thode_73}. {Therefore, in this work we do not consider issues like the wave coherency orthogonal to the wavevector or effects introduced by the beam front, which are the subject of future investigations}. Referring to Fig. \ref{Fig1}, beam 1 is {composed} of the
protons that have just been reflected. This beam moves into the upstream 
plasma, is rotated by the upstream magnetic field $\vec{B}$ and 
returns to the shock as beam 2 \cite{McClements_97}. 
The background electrons and protons 
represent the upstream plasma prior to its shock encounter. In total, 
we thus have four plasma species, {which are all modelled in a fully
kinetic treatment due to the importance of the ion reaction to the
wave fields} \cite{Schamel_96}. We place our simulation box into the 
upstream plasma close to the shock. We neglect the fact that the reflection of protons due to the shock motion cannot produce perfectly anti-parallel beams, and align the mean 
velocity vectors of both beams with the simulation or {{\em x}}-direction. 
We also assume the plasma to be homogenous orthogonal to the beams. 
The mean speeds of the beams in the {{\em x}}-direction are $v_b$ for beam 1 and $-v_b$ 
for beam 2, so that the total current in the simulation box vanishes.
{Therefore, no return current develops for this initial condition.}
Initially, all components of the mean speed are zero for the background
electrons and protons, as are the mean speed components of the proton
beams orthogonal to the simulation box. We limit our simulation time to
much less than an inverse proton gyrofrequency, thereby eliminating the
effects arising from the magnetic rotation of the beam. The spatial 
scale of our simulation box is small compared to the foreshock region, so that we may 
take all four species as spatially homogeneous. 

As initial parameters for all simulations, we take the electron plasma 
frequency $\omega_{p,e} = {(n_e e^2 / \epsilon_0 m_e)}^{1/2}=2\pi 
\times 10^5 s^{-1}$. Here $n_e$, $e$, $\epsilon_0$ and $m_e$ denote the 
electron number density, the magnitude of the elementary charge, the 
dielectric constant and the electron mass, respectively. The number 
densities of beam 1 and beam 2 are $n_{b1} = n_{b2} = n_b = n_e / 10$ 
in the rest frame of the respective beams. {Such densities for the shock
reflected ion beams are representative for perpendicular shocks}
\cite{Lembege_04}. The density of the background 
protons is $n_p = n_e - 2\gamma (v_b) n_b$, such that the net charge in the 
simulation box vanishes. The initial thermal spread of the electrons is 
$v_{th,e} = {(K_b T_e / m_e)}^{1/2} = 10^{-2}c$. 

For the simulations of beam
instabilities in an unmagnetized plasma in this section, the background protons and beam 1 have the thermal 
speed $v_{th,p} = v_{th,b1} = \sqrt{10}v_{th,e} {(m_e / m_p)}^{1/2}$, 
where $m_p$ is the proton mass. In all simulations, we use $m_p / m_e
= 1836$. Proton beam 2 has the thermal speed 
$v_{th,b2}=10 v_{th,b1}$. The proton temperature is increased relative 
to the electron temperature to improve the resolution of the proton 
velocity distributions by the Vlasov code. For the simulations in a
magnetized plasma, to be discussed in the next section, we use the 
values $v_{th,p} = v_{th,b1} = v_{th,e} {(m_e / m_p)}^{1/2}$ and 
$v_{th,b2}=10 v_{th,b1}$.

All particle species have a thermal speed that is small compared to the 
considered beam speeds $v_b = 0.6c$ (slow beams) and $v_b = 0.9c$ 
(fast beams), and are thus cold. The temperatures of the background 
electrons and protons are higher than a few eV, and their number 
densities are lower than $\approx 10^{10} m^{-3}$, estimated for 
the accretion disk of an AGN \cite{Jones_01}. Our choice of initial 
conditions is due to restrictions in the computer time.
No accurate phase space distribution functions are available for the 
beams at relativistic shocks and we thus take them to be cool, as in 
Ref. \cite{Dieckmann_04b}. In this section, our plasma is 
unmagnetized so that the developing instability is that discussed by Buneman in
Ref. \cite{Buneman_58}, except for a factor of $\gamma (v_b) = 
1/(1-v_b^2 / c^2)$, which has been examined numerically in Ref.
\cite{Thode_73}. 
To a good approximation, the most unstable Buneman wave is expected to grow at the wavenumber $k_u = \omega_{p,e}/v_b$, at a frequency $\omega_u = 
\omega_{p,e}$, and with the linear growth rate 

\begin{equation}
\omega_{im} = \gamma^{-1} (v_b){(3\sqrt{3} \omega_{p,b}^2 \omega_{p,1}/16)}^{1/3},
\label{growthRate}
\end{equation}
where $\omega_{p,b}$ is the beam plasma frequency.

We use $M=512$ grid cells 
along the {{\em x}}-direction for both the Vlasov and PIC simulations. The total 
box length is $L = 2\lambda_u = 4\pi / k_u$, and we normalize time by 
$T_p = 2\pi / \omega_{p,e}$. We analyse the electric field calculated by the simulations by first 
taking a Fourier transform over space and time. We separate the 
$\omega,k$ spectrum into two sets of two quadrants that have either 
$\omega / k < 0$ or $\omega / k > 0$. The consecutive inverse Fourier 
transforms applied to both data sets reconstruct the wave fields for 
the positive and negative phase speeds separately. In the absence of 
nonlinear mode coupling, we would thus separate the wave fields driven 
by both proton beams. We obtain the field amplitudes as a function of 
$k, t$ by applying 
\begin{equation}
E(k_j,t) = M^{-1} \left | \sum_{l=1}^{M} E_x (x_l, t) \exp{(-ik_jx_l)}
\right |
\label{Eq1}
\end{equation}
to the electric field. The amplitudes of the electrostatic Buneman waves 
are given in physical units V/m. 

\subsection{Slow beam}

\begin{figure}
	\begin{center}
	\epsfxsize=8.8cm
	\epsfbox{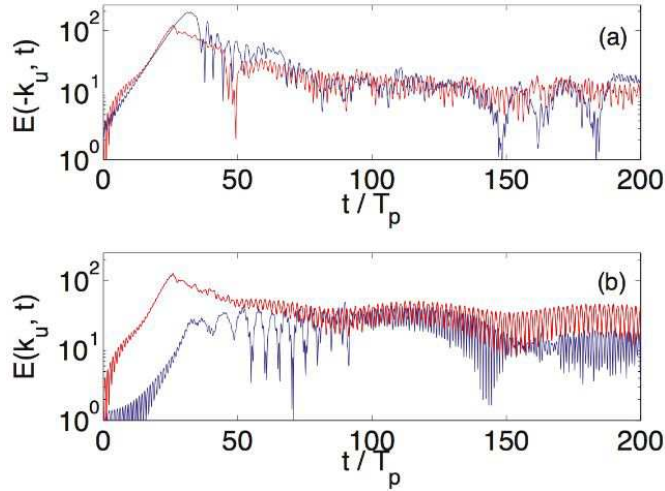}
	\end{center}
	\caption{\label{Fig2} Comparison of waves driven by proton beams with $v_b = 0.6c$. 
The logarithmic amplitudes of the waves with a negative phase speed ($k_u < 0$) 
are shown in (a) and with $k_u > 0$ in (b). PIC results are shown in blue and Vlasov results in red.}
	\end{figure}

The PIC simulation for $v_b=0.6c$ uses 1600 particles per cell (ppc) for the electrons 
and 784 ppc for each of the proton species. The Vlasov simulation uses 
8192 equidistant cells in $p_x$ covering the momentum range $-4.4\, < \, (p_x / m_e c)\, < \, 4.4$.
For both simulations we calculate the amplitudes $E(k_u,t)$ and $E(-k_u,t)$ from the computed $E_x$ using equation \ref{Eq1}  and compare the results in figure \ref{Fig2}.

The electric field in both simulations grows initially exponentially.
The experimentally measured growth rate  in the PIC simulation is $\Gamma_{ex} \approx 
0.97 \omega_{im}$, with $\omega_{im} / 
\omega_{p,e}=0.0228$ inferred from Eq. \ref{growthRate}. In the Vlasov 
simulation the measured growth rate is practically identical. 
We observe from figure \ref{Fig2} that both waves in the Vlasov 
simulation grow symmetrically, whereas the wave with $\omega_u / 
k_u < 0$ in the PIC simulation grows earlier, and to a larger peak 
amplitude.

Vlasov codes are free of noise, so we introduce an initial perturbation 
to encourage the growth of instabilities. We apply the initial perturbation 
to both beams in the Vlasov simulation at $k_u$ as in \cite{Luque_05}, 
and therefore the initial electric field amplitudes for both 
waves are identical. The thermal spread of both beams and their
difference is small compared to $v_b$, and the growth rates of the 
instabilities driven by both beams are thus similar. The two waves in 
the Vlasov simulation thus grow at the same rate to a comparable 
saturation amplitude, as shown in figure \ref{Fig2}. 

In the PIC simulation, on the other hand, the initial wave amplitudes
are given by noise, and the noise levels are higher for the hotter beam
\cite{Dieckmann_04d}. Therefore, the wave with $\omega / k < 0$ grows 
first. Since the weak wave with positive $\omega_u / k_u$ in the PIC 
simulation can interact nonlinearly with a limited phase space interval 
compared to that of the waves in the Vlasov simulation, the wave with 
negative $\omega_u / k_u$ can grow to a larger amplitude without 
interacting with the second wave, {e.g. by the formation of stochastic
bands due to the simultaneous interaction of electrons with both waves} 
\cite{Escande_82}.

\begin{figure}
\begin{center}
\epsfxsize=8.8cm
\epsfbox{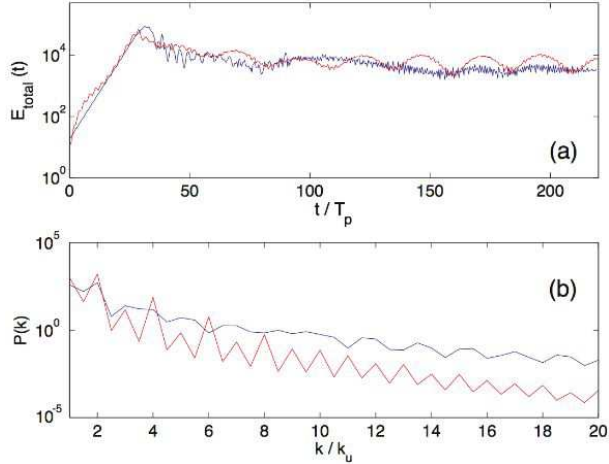}
\end{center}
\caption{\label{Fig3}Logarithmic plots of the electrostatic field energy density and the
{{\em k}}-spectrum for $v_b = 0.6c$. (a) The total electrostatic 
field energy density. (b) The power spectrum, integrated 
from $t_1$ to $t_2$, of the total electrostatic 
field as a function of $ k / k_u $. Results from the PIC simulation are given in blue and from the 
Vlasov simulation in red.}
\end{figure}

The electric fields in both simulations saturate by the trapping of 
electrons as discussed for almost identical parameters in {Ref.} 
\cite{Dieckmann_04b} and where the term trapping is used to denote 
electron reflection by the wave potential, as in \cite{Bauer92}. 
As a result, phase space holes form in the electron distribution 
\cite{Luque_05,Schamel_86,Eliasson_05,Bernstein_57}. The islands of trapped 
electrons have only a finite lifetime at this $v_b$ \cite{Schamel_96} 
and collapse by the sideband instability \cite{Kruer_69,Krasovsky_94}. 
Both simulations show a comparable lifetime for the nonlinearly 
saturated electrostatic wave. After $t\approx 70 T_p$, the electric field amplitude fluctuates around a low and constant mean 
value. Further evidence for similar development of the electrostatic instability
in both codes is provided by the energy density of the 
electrostatic field $E_{total} = M^{-1} \sum_{j=1}^{M} E^{2}(x_j,t)$ 
and the wave spectrum $P(k) = {(t_2-t_1)}^{-1}\int_{t1}^{t2}
E^2 (k,t) dt$, with $t_1 =200T_p$ and $t_2=220 T_p$, which are shown in 
figure \ref{Fig3}.
The energy density of the electric field peaks at $t\approx 30 T_p$
and decreases thereafter, primarily because of the collapse of the
electrostatic waves. At the end of the simulation, both codes show a comparable
{{\em k}}-spectrum at low wavenumbers. The PIC code gives stronger wave
power at higher wave numbers, primarily due to its higher noise levels.

\begin{figure}
\begin{center}
\epsfxsize=8.8cm
\epsfbox{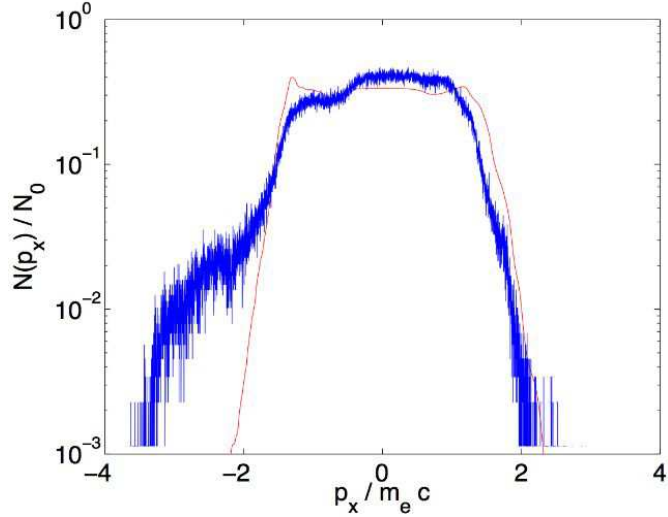}
\end{center}
\caption{\label{Fig4}The electron momentum distribution for
$v_b = 0.6c$ at $t=220T_p$ in the PIC simulation (blue)
and in the Vlasov simulation (red).}
\end{figure}

The nonlinear interaction of the electrons with the electrostatic
waves has increased the electron energy. Both simulation codes show, 
at $t=220 T_p$, a similar electron momentum distribution for $p_x>0$, see Fig.\ref{Fig4}. 
For $p_x<0$ the electron momentum distribution calculated
by the PIC code shows a tail.
The higher peak momentum at negative $p_x$ in the PIC simulation
is due to the larger amplitude of the wave with $k=-k_u$ compared
to that in the Vlasov simulation, as shown in figure \ref{Fig2}. The
larger amplitude leads to an island of trapped electrons that is
wider in $p_x$; the collapse of the stronger wave thus redistributes 
the electrons over a wider $p_x$ interval. The weaker wave with 
$k=k_u$ in the PIC simulation cannot accelerate electrons to the 
same peak momentum as the waves in the Vlasov simulation, because initially
it only accesses a smaller $p_x$ interval.
Conversely, the equally strong electrostatic waves with $k=k_u$ and $k=-k_u$ 
in the Vlasov simulation lead to the observed symmetric momentum 
distribution.

\subsection{Fast beam}

\begin{figure}
\begin{center}
\epsfxsize=8.8cm
\epsfbox{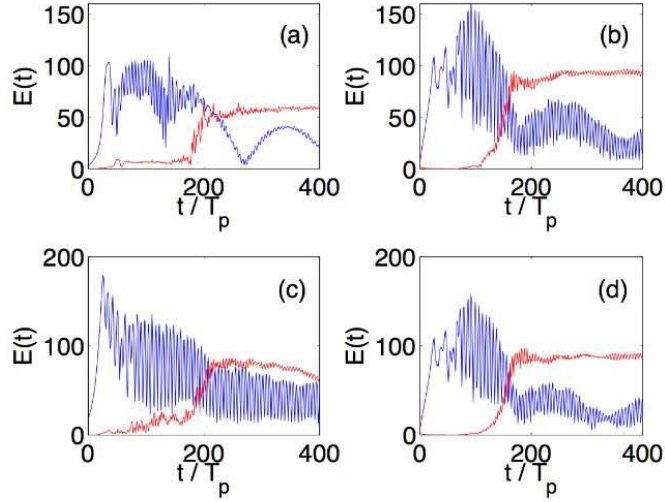}
\end{center}
\caption{\label{Fig5}Waves driven by the beams with $v_b = 0.9c$. The blue traces correspond to the waves with $k=k_u$ and the red to the waves with $k=k_u/2$. 
The amplitudes of the waves with positive phase speed ($k > 0$) 
are shown in (a) for the PIC simulation and in (b) for the Vlasov 
simulation. The amplitudes of the waves with negative phase speed 
($k < 0$) are shown in (c) for the PIC simulation and in (d) for 
the Vlasov simulation.}
\end{figure}

The PIC simulation for $v_b=0.9c$ uses 3200 particles per cell (ppc) for the electrons and 2048 ppc 
for each of the proton species. The Vlasov simulation uses 8192 
equidistant cells in $p_x$ covering the momentum range from 
$-20\, < \, (p_x / m_e c)\, < \, 20$. We show the amplitudes 
$E(k_u,t)$ and $E(k_u/2,t)$ in figure \ref{Fig5}.

\begin{figure}
\begin{center}
\epsfxsize = 8.8cm
\epsfbox{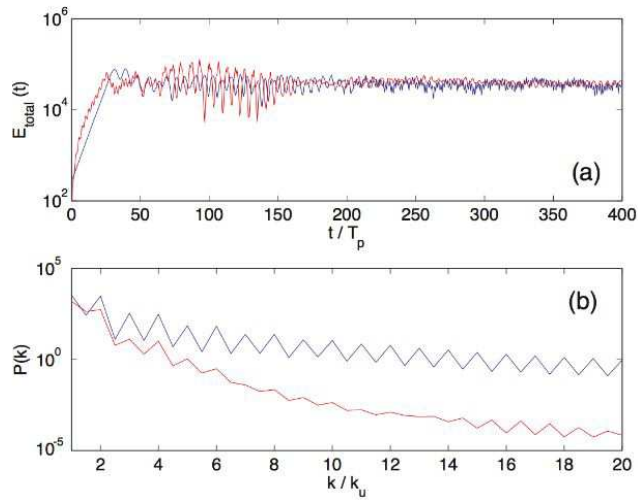}
\end{center}
\caption{\label{Fig6}Logarithmic plots of the electrostatic field energy density and k-spectrum 
for $v_b = 0.9c$ in the PIC simulation (blue) and in the Vlasov simulation (red). (a) The total electrostatic field energy density (b) The power spectrum, integrated from $380T_p$ to $400T_p$, of the total electrostatic field as a function of $ k / k_u $.}
\end{figure}

\begin{figure}
\begin{center}
\epsfxsize = 8.8cm
\epsfbox{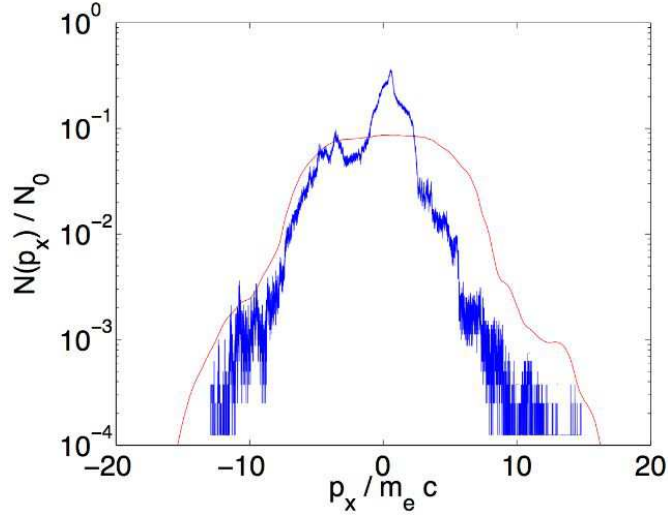}
\end{center}
\caption{\label{Fig7}The electron momentum distribution for
$v_b = 0.9c$  at $t=400T_p$ in the PIC simulation 
(blue) and the Vlasov simulation (red).}
\end{figure}

The anticipated growth rate $\omega_{im} \approx 0.0152$, inferred from Eq. \ref{growthRate}, is
accurately reproduced by the PIC simulation. In the Vlasov 
simulation, the maximum growth rate is $\Gamma_{ex} \approx 0.78
\omega_{im}$.
From figure \ref{Fig5} we find that initially the wave with $k=k_u$, which is driven initially by the Buneman instability, is
dominant, but at $t\approx 200 T_p$ its amplitude decreases
and a wave grows at $k_u / 2$. The coupling of the wave
energy towards lower wavenumbers is a characteristic of the
sideband instability \cite{Krasovsky_94}, and the decrease in the
amplitude of the $k=k_u$ wave is associated with the collapse or merger 
of islands of trapped electrons, as discussed in Refs. 
\cite{Dieckmann_04b,Schwarzmeier_79}.
The time lag between the initial nonlinear saturation and the coupling of energy towards the sideband modes thus provides a measure of the non-linear stability of the electrostatic instability (which in the initial, linear phase is governed by the Buneman waves). Both codes show here a lifetime of about $150 T_p$. Figure \ref{Fig6} shows that the energy density in the electrostatic field as a function of
time is almost identical in both 
codes; so is the wave power, integrated between $t_1=380 T_p$ and 
$t_2 = 400 T_p$, at low wavenumbers. Again, the PIC code
shows higher noise levels at larger wavenumbers. 
The electron momentum distributions at the end of the PIC and 
Vlasov simulations agree reasonably well, as shown in figure \ref{Fig7}.
The electrons in both simulations reach a peak momentum
$|p_x| \approx 15 m_e c$, however the Vlasov simulation has a flatter central 
maximum. The PIC code also evolves to a flat-top distribution, but 
more slowly.

\subsection{Particle distributions at late-time}
\begin{figure}
\begin{center}
\epsfxsize = 12cm
\epsfbox{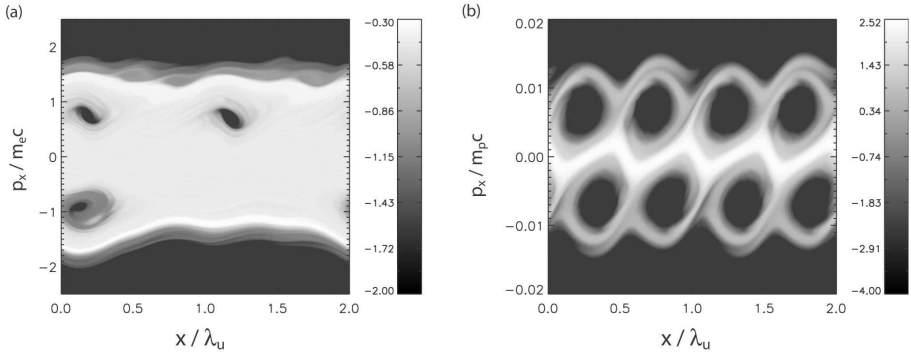}
\end{center}
\caption{\label{Fig8}The final electron (a) and bulk proton (b) phase space distributions for
$v_b = 0.6c$  at $t\approx 200T_p$ in the Vlasov simulation. The colour axis
denotes the logarithmic magnitude of the distribution functions, $f_e$ and $f_i$.}
\end{figure}

 We note that the PIC work in Ref. \cite{Dieckmann_04b}  yields a flat-top 
distribution at the later time $t \approx 1000 T_p$, whose plateau has a width $\approx 6 m_e c$ {and which shows an electron phase
space distribution similar to that of the Vlasov simulation shown
in Fig. \ref{Fig8}. The proton beam distributions computed by the Vlasov
simulation, for the case of $v_b = 0.6c$ display ion holes which resemble qualitatively those in Ref. \cite{Dieckmann_04b}. However, in the case of the Vlasov simulations presented here, ion holes develop almost symmetrically with both positive and negative phase {speeds},
rather than the `one sided' distribution observed in PIC simulations. This is shown in Fig.\ref{Fig8}b for the bulk protons and is attributable to the comparable amplitudes for instabilities associated with both beams. The two strong electrostatic waves can each decay into a low frequency mode with $k=2k_u$ and a second Langmuir wave. In the PIC case, the ESWs associated with beam 2 develop first, as discussed earlier.} 
In both the slow and fast beam cases, and for both Vlasov and PIC codes, the proton distributions at late time are non thermal. In the absence of a magnetic field, their continuing evolution will be governed primarily by the effective collisionality, which naturally results from the finite resolution of the numerical schemes \cite{Schamel_96,Korn_96}, {from particle collisions with the fluctuating electric fields \cite{Morales_74} and the stochastic interaction of the electrons with the electric fields of the ion beam structures} \cite{Escande_82}. Hence, the time asymptotic equilibrium {shown in figure 8} is similar to the structural dissipative equilibria described in {the Refs.} \cite{Schamel_96,Korn_96}.

\section{Magnetic field effects}

In the preceeding section, we established that the beam-excited electrostatic waves can remain non-linearly stable for one hundred plasma oscillation periods or more.
We have shown that {the results from} PIC and Vlasov simulations agree on this, which is evidence for a physical origin behind the wave stabilization. In the present section, we assess the efficiency
of these waves with respect to trapping and ESA: this requires the inclusion of an {external} magnetic field. Under these conditions, our computational approach is based on PIC simulations. We use 512 simulation cells to resolve the {{\em x}}-direction of the simulation box
with its length $L = 4\pi / k_u$. The minimum simulation
wavenumber $2\pi / L < k_u$ allows for the development of 
the sideband instability and the merger of trapped particle islands, 
and thus corresponds to the case
of the long simulation box in Ref. \cite{McClements_01}. 
The magnetic field orientation is aligned with the {{\em z}}-axis and 
this orientation, together with the relativistic wave speeds,
makes the results directly comparable to the work in Ref.
\cite{Katsouleas_83}.

We choose values for the magnetic field strength that correspond to an 
electron gyrofrequency $\omega_{c,e}=\omega_{p,e}/10$
for the simulation with $v_b = 0.6c$, and 
$\omega_{c,e}=\omega_{p,e}/10$ and $\omega_{c,e}=\omega_{p,e}/100$
for the two simulations with $v_b = 0.9c$. With these ratios we can 
resolve all relevant electron timescales without requiring
long simulation times. The weak magnetic field equals that in
Ref. \cite{McClements_01} and the strong magnetic field 
that in Ref. \cite{Dieckmann_99}. In both these works, the 
authors considered nonrelativistic beam speeds. 

Typical ratios for $\omega_{p,e} / \omega_{c,e}$ may be larger in 
astrophysical environments. We get, for example, the ratio  
$\omega_{p,e}/\omega_{c,e} =3200$ if we use the 
estimated average magnetic field $B = 10$ nT and the estimated
average electron number density $n_e = 10^{10} m^{-3}$ for 
the accretion disk orbiting the central black hole in NGC 4261 
\cite{Jones_01}. Spatial and temporal fluctuations in the values 
of $B$ and $n_e$ could, however, lead to 
lower local characteristic frequency ratios in the accretion disk.

\subsection{Slower mildly relativistic beam, $v_b = 0.6 c$}

\begin{figure}
\begin{center}
\epsfxsize=8.8cm
\epsfbox{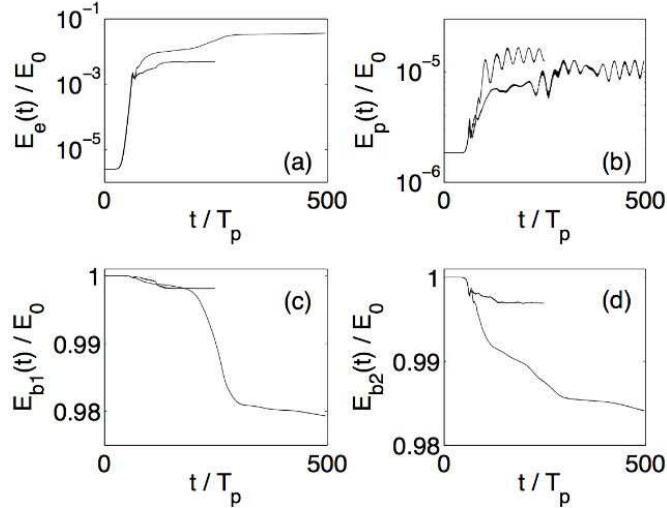}
\end{center}
\caption{\label{Fig9} Time evolution of the total energy of different particle species in
units of the initial energy of proton beam 1, $E_0$, for $v_b = 0.6c$. The 
short traces correspond to unmagnetized plasma, long traces to  magnetized plasma. (a) Logarithm of the electron kinetic energies;
(b) logarithm of the kinetic energy of the background protons; (c) shows the 
energy of proton beam 1; (d) the energy of proton beam 2.}
\end{figure}

We resolve the electron distribution by 1600 ppc and each of the 
proton species by 784 ppc. The simulation duration is $480 T_p$ and
during this time the proton beams rotate by an angle 
$\alpha \approx 7.5^o$. The electric field follows closely that of 
the unmagnetized plasma as shown in figure \ref{Fig2}. Figure \ref{Fig9} depicts the time evolution 
of the kinetic energies of the plasma species for unmagnetized and magnetized regimes. The 
electrons are accelerated most rapidly during the initial linear 
stage of the developing instability ($t < 50 T_p$), when they are 
first trapped by the wave. Here, the electron acceleration is almost 
identical in both magnetised and unmagnetised cases. This is also 
reflected by the similar energy loss of the beams to the background 
electrons and protons during this time. At $t \approx 50 T_p$ the 
curves diverge. The particle energies of all species in an 
unmagnetized plasma reach a plateau, i.e. the instability has been 
quenched. Proton beam 2, which has been connected to the stronger 
wave in figure \ref{Fig2}, has lost more energy. 
In the magnetized plasma, on the other hand, the electron energy
continues to rise after $t \approx 50 T_p$. The electron energy
is here a few percent of the beam energy, and the continuing acceleration
extracts noticeable fractions of the beam energy. The ESA extracts
energy first from proton beam 2 because the associated strong wave traps
electrons first. In contrast, the energy of proton beam 1 drops rapidly
at around $t \approx 250 T_p$. The energy gain by the background protons is
comparable for the unmagnetized and magnetized plasmas.

\begin{figure}
\begin{center}
\epsfxsize=8.8cm
\epsfbox{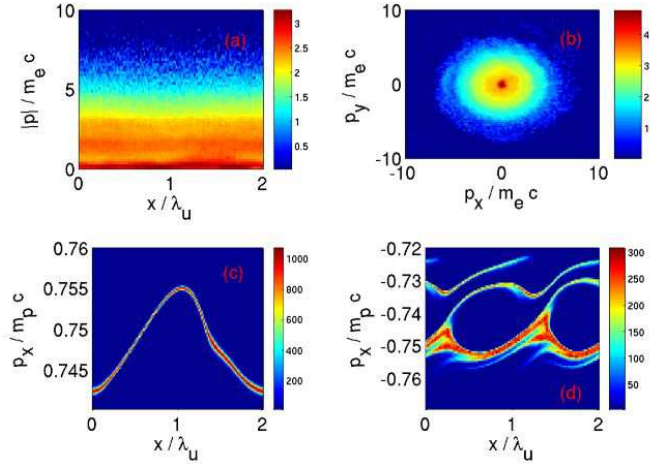}
\end{center}
\caption{\label{Fig10} Particle distribution functions at $t = 207 T_p$
for $v_b = 0.6c$. All phase space densities reflected by the colour scale are in units of the $\log_{10}$ of the number of simulation particles. (a) The spatially homogeneous electron 
momentum distribution; (b) the hot thermalized electron momentum 
distribution; (c) the strongly oscillating proton beam 1; (d) well-developed phase space holes in the distribution of proton beam 2.}
\end{figure}

During the initial stage, beam 2 is dominant in the ESA. The 
plasma distribution at $t = 207 T_p$ is shown in figure \ref{Fig10}.
The electrons have reached a spatially isotropic and gyrotropic 
momentum distribution. We discuss below the mechanism by which the electrons 
have been thermalized orthogonally to $\vec{B}$, for the 
simulations with $v_b = 0.9c$.
We find strong momentum oscillations of beam 1 and well-developed 
phase space holes in beam 2. These beam distributions are further 
evidence of a much more strongly nonlinear wave associated with beam 2. 

\begin{figure}
\begin{center}
\epsfxsize=8.8cm
\epsfbox{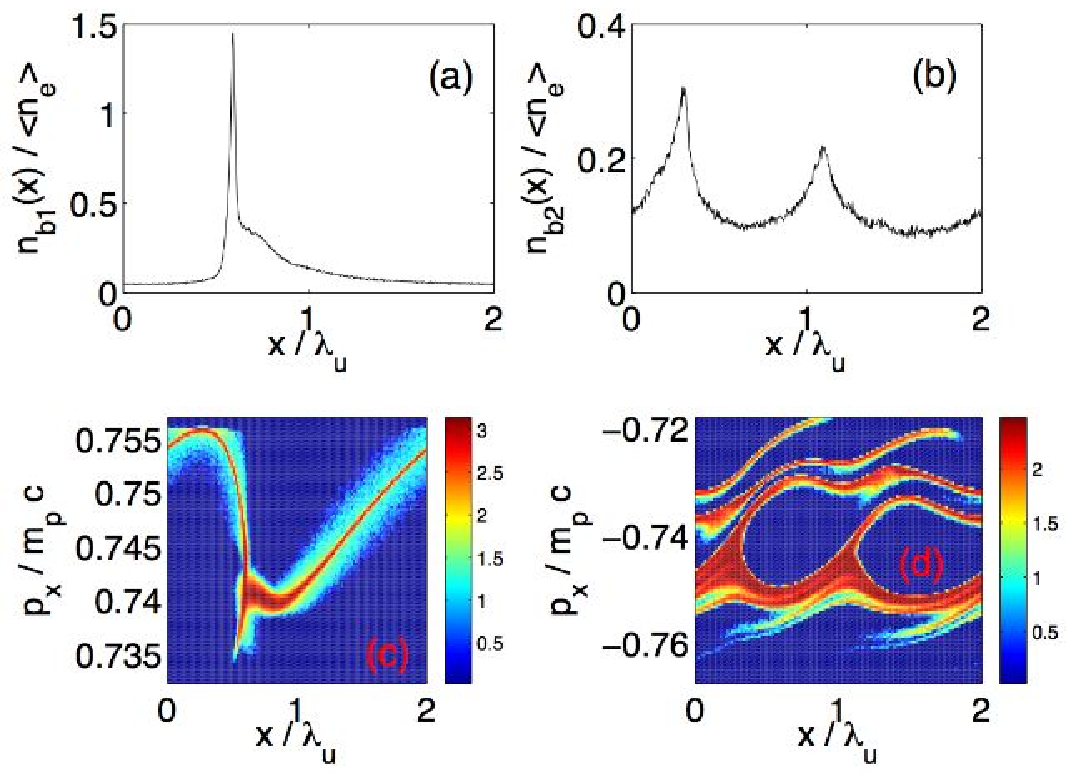}
\end{center}
\caption{\label{Fig11}Proton beam distribution functions at $t = 250 T_p$
for $v_b = 0.6c$:
(a) and (b) show the number density of beam 1 and beam 2, respectively, 
in the box frame of Ref. in units of the background electron number density;
(c) and (d) show the phase space distributions of beams 1 and 2 respectively. The colour scale is in units of the $\log_{10}$ of the number of simulation protons.}
\end{figure}

The rapid drop in the energy of beam 1 is associated with spatial 
concentration leading to the development of a density peak shown 
in figure \ref{Fig11}(a) where the local beam density exceeds the 
average density of the background electrons. {Strong oscillations at $2\lambda_u$ produce a density modulation which leads to non-linear steepening of wave-fronts as they traverse the system. This results in the observed density spike. The corresponding ``bunching'' in phase space indicates the onset of wave breaking and is accompanied by a jump in the mean speed of 
the beam, see figure \ref{Fig11}(c)}. 
{The development of this density
spike is similar to that forming at the leading edge of plasma
expanding into vacuum. This is detailed in 
Ref. \cite{Sack_1987} where it is discussed analytically, using a hydrodynamic
self-similar and non-relativistic approach, how a sharp ion front 
develops in response to a preceding electron cloud. In such a
system, the decreasing velocity profile in the expansion direction
of a compressional wave, that expands into a less dense medium 
implies a steepening and eventually breaking of the wave. At the point of wave-breaking, the distribution function becomes multi-valued and the fluid approximation breaks down, but a treatment based on the Vlasov equation (also detailed in Ref. \cite{Sack_1987}) demonstrates the formation of structures qualitatively similar to those displayed in Fig.\ref{Fig11}(c). Furthermore, beam momentum jumps associated with a 
strong beam charge density modulation, have also been reported in 
Ref. \cite{Gohda_04}.}

\begin{figure}
\begin{center}
\epsfxsize=8.8cm
\epsfbox{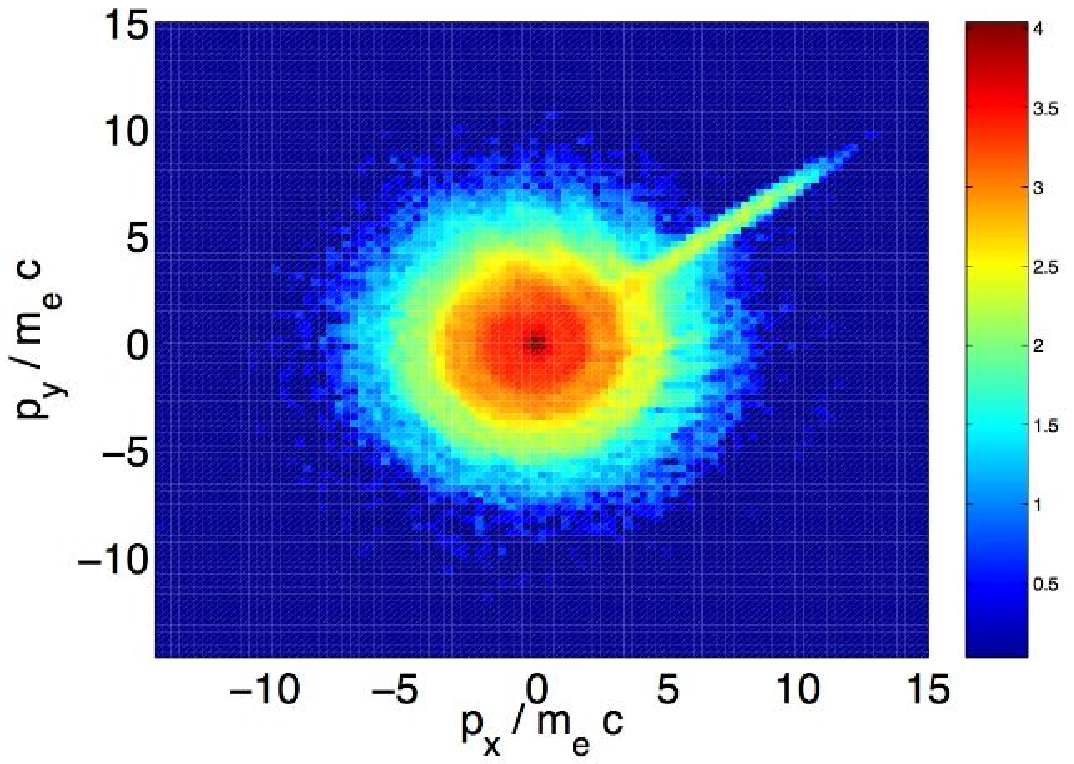}
\end{center}
\caption{\label{Fig12} The electron momentum distribution at $t = 250 T_p$
for $v_b = 0.6c$: The colour scale is in units of the $\log_{10}$ of the number of simulation electrons.}
\end{figure}

The difference between figure \ref{Fig11}(c) at $t=250T_p$ and its earlier counterpart figure \ref{Fig10}(c) at $t=207T_p$ reflects, together with \ref{Fig11}(a), the key physical process underlying localised ESA in the present context. Its consequences for the electron acceleration are immediately visible in figure \ref{Fig12} at $t=250T_p$, which is the counterpart of the earlier figure \ref{Fig10}(b) at $t=207T_p$. The local density accumulation of beam 1 is connected to a strong
electrostatic wave potential moving towards increasing values of {{\em x}}.
The resulting electric field is strong enough to trap electrons,
but only over the spatially localised high density peak. The result is a localised
ESA which leads to an electron beam that accelerates out of the
heated electron distribution. This electron beam and the associated
``frying-pan'' distribution are shown in figure \ref{Fig12}.

\begin{figure}
\begin{center}
\epsfxsize=8.8cm
\epsfbox{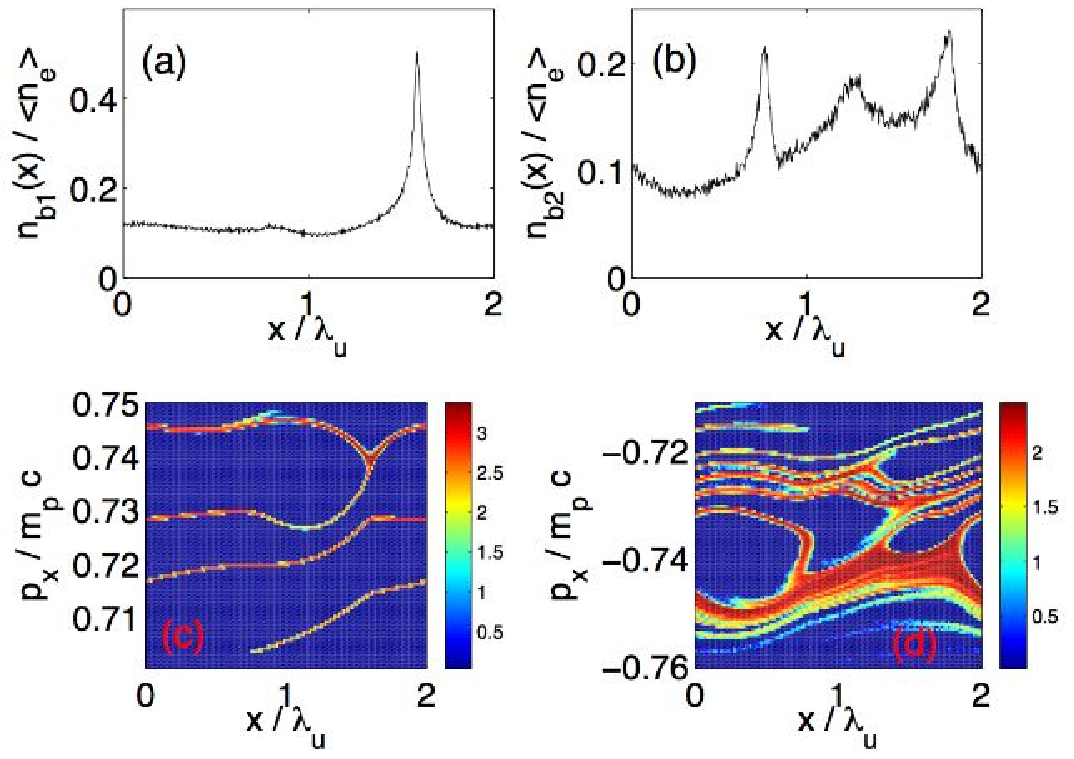}
\end{center}
\caption{\label{Fig13} Proton beam distribution functions at $t = 455 T_p$
for $v_b = 0.6c$:
(a) and (b) show the number density of beam 1 and beam 2 respectively in 
the box frame of Ref., in units of the electron number density;
(c) and (d) show the phase space distribution of beams 1 and 2, respectively. 
Both colour scales are in units of the $\log_{10}$ of the number of computational protons.}
\end{figure}

The local electron acceleration counteracts further proton beam charge
accumulation and eventually reduces the beam potential, so that the
electrons detrap. This detrapping arrests both the electron acceleration
and the associated energy loss of proton beam 1 in figure \ref{Fig9} at $t \approx 270 T_p$.
The beam density peak evolves to the much less pronounced but still
strongly localized spike shown in figure \ref{Fig13}. At this time we 
also observe the onset of a merger of the two phase space holes in 
proton beam 2.

\begin{figure}
\begin{center}
\epsfxsize=8.8cm
\epsfbox{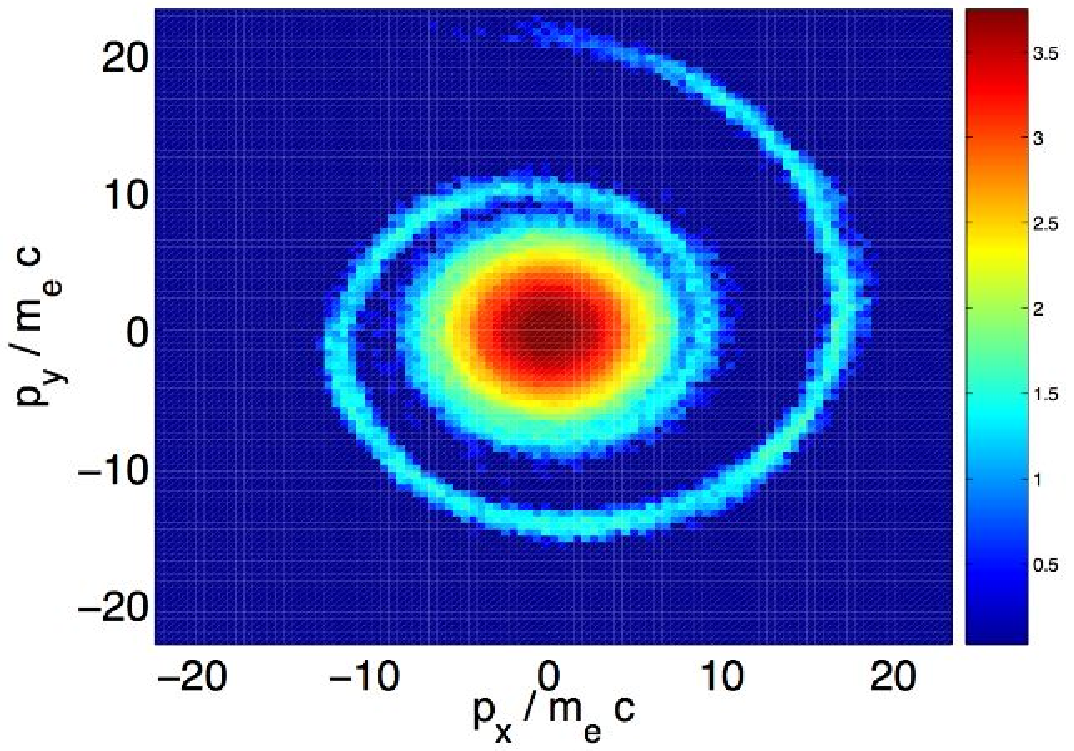}
\end{center}
\caption{\label{Fig14} The electron momentum distribution at 
$t = 455 T_p$ for $v_b = 0.6c$: The colour scale is in units 
of the $\log_{10}$ of the computational electrons. }
\end{figure}

The electrons from the handle of the ``frying-pan'' distribution shown
in figure \ref{Fig12} have detrapped at this time, and they gyrate freely 
in the magnetic field. Their relativistic mass increase reduces their 
rotation frequency for increasing momenta, and the accelerated beam electron distribution therefore develops into the spiral shown in figure \ref{Fig14}. 

\subsection{Faster mildly relativistic beam, $v_b = 0.9c$}

Due to the anticipated larger momentum range accessible to the electrons
we use 3200 ppc for the electrons and 2048 ppc for the protons in our
PIC simulations, which provides adequate momentum space resolution. 
We model the plasma for a duration of $980T_p$. In the simulation with
$\omega_{p,e} /\omega_{c,e} = 10$ the proton beams have rotated by $8.4^o$.

\begin{figure}
\begin{center}
\epsfxsize=8.8cm
\epsfbox{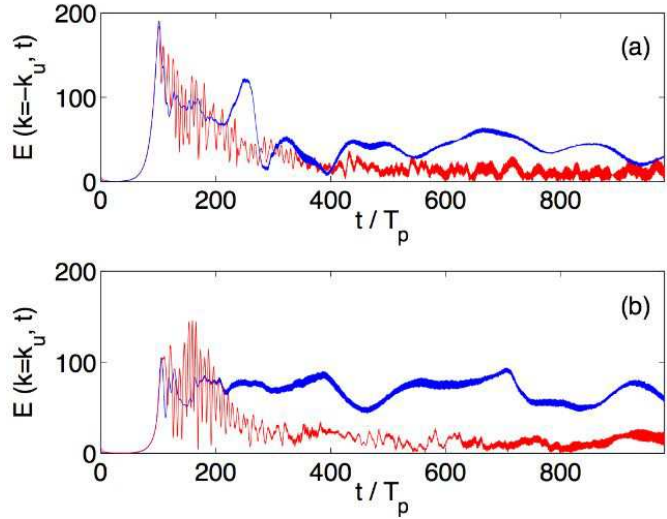}
\end{center}
\caption{\label{Fig15} Electrostatic waves driven by the proton beams with $v_b = 0.9c$. 
The blue trace corresponds to $\omega_{p,e} = 10 \omega_{c,e}$
and the red trace to $\omega_{p,e} = 100 \omega_{c,e}$.
The amplitudes of the waves with negative phase speed ($k_u < 0$) 
are shown in (a), while (b) shows the amplitudes of the waves with 
$k_u > 0$.}
\end{figure}

The electrostatic field in the weakly magnetic simulation with 
$\omega_{p,e}=100\omega_{c,e}$ (red curves in figure \ref{Fig15}) 
grows similarly to that in the unmagnetized plasma in figure \ref{Fig5}. 
The visible rapid fluctuations are similar to those that have been 
attributed to the trapping of electrons \cite{Dieckmann_04b}. The wave 
in the more strongly magnetised simulation with 
$\omega_{p,e}=10\omega_{c,e}$ (blue curves in figure \ref{Fig15}) does not 
show these.
This is possibly due to rapid electron acceleration by the magnetic
field. The electrons gain substantial energy, and thus relativistic mass, 
during a bounce period in the wave potential. The relativistic mass
in this case depends strongly on the initial phase space position of
the electron, and no coherent oscillations of trapped electrons occur.

\begin{figure}
\begin{center}
\epsfxsize=8.8cm
\epsfbox{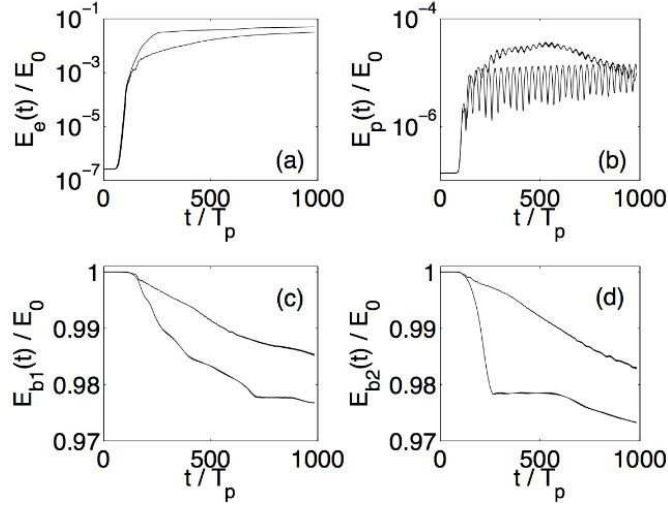}
\end{center}
\caption{\label{Fig16} The time evolution of the particle energies in
units of the initial energy of beam 1, $E_0$, for $v_b = 0.9c$ and for
the weak and strong magnetic fields. (a) Electron kinetic 
energies: the upper trace corresponds to the strong magnetic field. 
(b) kinetic energy of the background protons: the strongly 
oscillating lower trace corresponds to the weak magnetic field. (c) Energy of beam 1: the lower trace corresponds to the 
strong magnetic field. (d) Energy of beam 2: the lower 
trace corresponds to the strong magnetic field.}
\end{figure}

Figure \ref{Fig16} confirms the more rapid acceleration of electrons in the simulation with
$\omega_{p,e} = 10 \omega_{c,e}$ as follows. The curves representing the kinetic energies of the electrons evolve practically identically until $t\approx 125 T_p$. The initial acceleration 
mechanism is here, as in figure \ref{Fig9}, the trapping of electrons by the growing wave potential. 
After this time the electrons are accelerated more rapidly in the 
simulation with $\omega_{p,e} = 10 \omega_{c,e}$ and their energy is 
about one order of magnitude larger than in the simulation with 
$\omega_{p,e} = 100 \omega_{c,e}$. As the simulations approach the 
end time $t = 980 T_p$, both curves appear to converge.

The background protons are only weakly accelerated. The high 
frequency oscillation of the proton energy for $\omega_{p,e} = 100 
\omega_{c,e}$ is likely to be the proton reaction to the oscillation 
of the corresponding (red) electric field in figure \ref{Fig15}. These 
high-frequency oscillations are almost absent in the case of 
$\omega_{p,e} = 10 \omega_{c,e}$ which is also in accordance with the 
absent field fluctuations in figure \ref{Fig15}. Both beams lose 
about 2\% of their initial energy during their non-linear interaction 
with the background plasma. Of particular interest is the rapid drop 
of the energy of beam 2 at $t\approx 200t_p$.

\begin{figure}
\begin{center}
\epsfxsize=8.8cm
\epsfbox{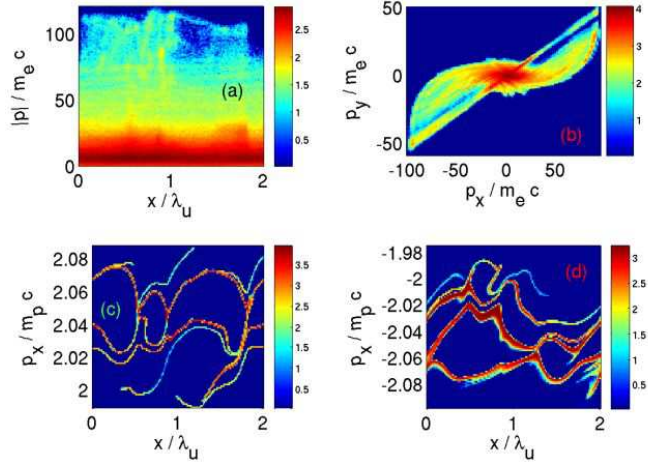}
\end{center}
\caption{\label{Fig17}Phase space distributions for $v_b = 0.9c$
and $\omega_{p,e} = 100 \omega_{c,e}$ at $t=980 T_p$.
(a) Spatially almost isotropic electron distribution
reaching a peak momentum in excess of $|p| = 100 m_e c$. (b)
Electron momentum distribution orthogonal to $\vec{B}$ is shown. (c) Phase space distribution of proton beam 1, (d)
that of beam 2.}
\end{figure}

The energy transfer from the beams for $\omega_{p,e} = 100 \omega_{c,e}$ accelerates the electrons to highly relativistic speeds. 
In figure \ref{Fig17} the well-defined phase space structure, with distributions that are
not yet spatially isotropic, is evidence that electrons 
are still trapped, and therefore undergoing ESA. This explains why electron energy is still increasing in figure \ref{Fig16}(a) at the end of the simulation. 

\begin{figure}
\begin{center}
\epsfxsize=8.8cm
\epsfbox{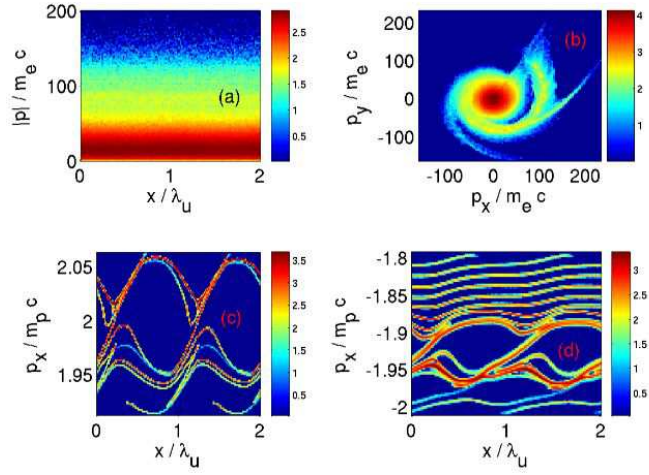}
\end{center}
\caption{\label{Fig18}The phase space distribution for $v_b = 0.9c$
and $\omega_{p,e} = 10 \omega_{c,e}$ at $t=980 T_p$:
In (a) we find a spatially isotropic electron distribution
reaching a peak momentum  $|p| = 180 m_e c$. In (b) the electron 
momentum distribution orthogonal to $\vec{B}$ is shown. In (c) we 
show the phase space distribution of proton beam 1, and in (d) that of 
beam 2.}
\end{figure}

For $\omega_{p,e} = 10 \omega_{c,e}$ the electron distribution extends to $|p| \approx 180m_ec$ and is spatially isotropic at the simulation's end, but not yet gyrotropic, 
as figure \ref{Fig18} shows. Both beams show phase space holes, which 
are the reason for the persistent electric field amplitudes (blue curves in figure 
\ref{Fig15}) even at late times.

The two attached movies show the time evolution of the electron
momentum space distribution in the $p_x,p_y$ plane for $v_b = 0.9c$
and for both magnetic field strengths. We observe the electrons along 
the {{\em x}}-direction towards increasing values of {{\em x}}; this has the effect of reversing the sense
of rotation relative to figures \ref{Fig17} and \ref{Fig18}. For weak magnetic field, 
Movie 1 shows the ESA for $\omega_{p,e}=100\omega_{c,e}$ and Movie 2 
shows the ESA for $\omega_{p,e}=10\omega_{c,e}$. In both movies, the 
total momentum is mapped into the colour scale. Movie 1 indicates that 
the phase space distribution remains unchanged except for a scaling 
factor that increases approximately linearly as a function of time.
Movie 2 shows that the strong magnetic field case yields a rapid acceleration
of electron beams which explains the rapid drop of the energy of beam 2
in the lower curve of figure \ref{Fig16}(d). It also shows repeated ESA bursts that
occur whenever the electric field amplitude, shown in blue in figure \ref{Fig15},
rises to a value at which it can trap electrons. Another important
observation from Movie 2 is the strong increase of the momentum spread
of the narrow electron beams that are produced by ESA. This beam heating
is most efficient when the electrons move parallel or antiparallel
to the x-axis. In this case, they can interact resonantly with the
electrostatic waves. This electron thermalisation resembles the
stochastic acceleration mechanism examined in Ref. \cite{Karney_77}
for nonrelativistic phase speeds of the waves.

\begin{figure}
\begin{center}
\epsfxsize=8.8cm
\epsfbox{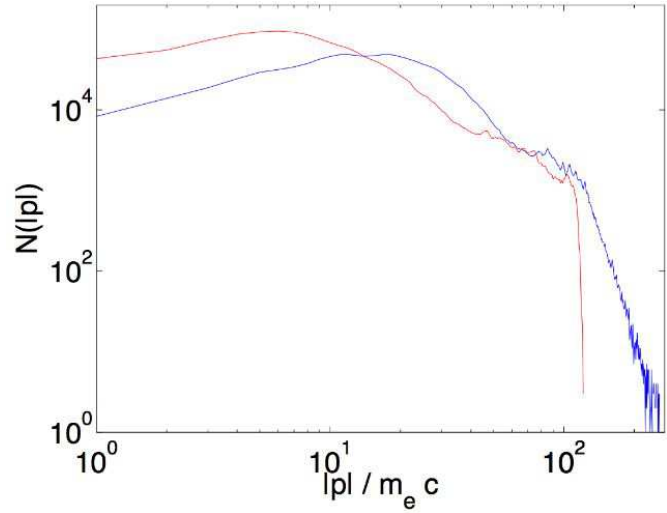}
\end{center}
\caption{\label{Fig19} The electron momentum density for beam speed $v_b = 0.9c$ at $t=980 T_p$:
The blue curve corresponds to the simulation with $\omega_{p,e}
=10 \omega_{c,e}$, and the red to $\omega_{p,e}
=100 \omega_{c,e}$. The density is given in units of simulation
electrons.}
\end{figure}

Finally, we integrate the electron distribution functions in figures
\ref{Fig17}(a) and \ref{Fig18}(a) over space to obtain their phase
space density as a function of the total momentum $|p|$, which we 
show in figure \ref{Fig19}.

Electrons in the more strongly magnetized plasma reach a
higher peak momentum. The heating in Movie 2 that, 
we believe, is due to stochastic heating, has apparently led to
the development of a power law distribution at high $|p|$.
In contrast the momentum distribution reached by the electrons in the more weakly
magnetized plasma (red trace in figure \ref{Fig19}) shows an abrupt
density drop at high $|p|$. Here, the fastest electrons are
still trapped and have a well-defined peak momentum. 

\section{Discussion}

In this paper, we have examined the efficiency of electron surfing
acceleration in producing relativistic electrons in plasmas. We have 
used both Vlasov and PIC computational approaches. The chosen initial conditions may be representative of the foreshock
region of plasma shocks in the accretion disks of microquasars, and 
of the central black holes of active galactic nuclei. We have found 
that across the beam velocity band between $v_b = 0.6c$ and $v_b=0.9c$, 
the peak momentum reached by the electrons increases by almost an order 
of magnitude. The most important variable is the increasing 
lifetime of the saturated electrostatic wave. This is also been observed 
in both the PIC and Vlasov codes, which yield quantitatively similar 
results. This similarity, despite the different methods by which 
both codes solve the relativistic Vlasov-Maxwell equations, is
evidence for a physical and not numerical origin for the increasing
lifetime of the saturated electrostatic wave. {This lifetime is restricted
for fast but non-relativistic phase speeds of the electrostatic waves}
\cite{Schamel_96,Korn_96}. In addition, the increasing phase speed of 
the wave, and thus the increasing Lorentz force that acts on the electrons, 
also contributes to stronger acceleration. The relation between the peak 
momentum reached by the electrons and the beam speed is thus a function 
that grows faster than linear. This trend persists up to $v_b = 0.99c$, 
for which the peak electron momentum has been shown to increase by another 
order of magnitude \cite{Dieckmann_04a}. ESA cannot however, accelerate 
electrons to a peak momentum exceeding $m_p / m_e$ \cite{Dieckmann_04a}.

We have compared the final momentum distributions for both beam speeds, 
for unmagnetized and magnetized plasmas. The introduction of a magnetic 
field with the ratios $\omega_{p,e} / \omega_{c,e}=$ 100 and 10
has increased the peak momentum reached by electrons, compared to
the case of unmagnetized plasma, from 
$|p|/m_ec\approx 2.5$ to $|p| / m_e c \approx 10$ for $v_b=0.6c$ and 
from $|p|/m_e c\approx 15$ to $|p|/m_ec \approx 200$ for $v_b =0.9c$.

We conclude that ESA works best if, first, the electrostatic waves
have a relativistic phase speed and, second, the magnetic field is strong
enough to accelerate the electrons significantly during the lifetime
of the saturated wave. In addition our simulations show that the
overall electron acceleration can be enhanced by secondary instabilities
that are triggered as a reaction to the strong electron acceleration.
We have found that the strong wave fields can lead to the local 
accumulation of beam protons which can, by their associated electric 
fields, yield further localised acceleration of electrons. {Such an 
accumulation of the beam density has, in other contexts, been investigated 
analytically for the case of a hydrodynamic plasma, expanding into a vacuum,
\cite{Sack_1987} and for a Q-machine experiment \cite{Gohda_04}}, and it 
may play an important part in electron acceleration to high energies in 
astrophysical environments. This spatially localized acceleration is 
responsible for the observed development of a frying-pan distribution 
that allows some electrons 
to double their momentum compared to that reached after their interaction 
with the Buneman wave. An inverse power law distribution can arise for 
the most strongly accelerated electrons.

The electron acceleration times, which are of the order of a few hundred 
plasma periods $T_p$, can yield relativistic electrons from a thermal 
pool in short times. ESA is thus a potential mechanism that could give 
rise to the rapid fluctuations in the emission intensity of radio 
synchrotron radiation from the accretion disks of microquasars. As an
illustration we can take the average density of $10^{10}/m^3$ reported
for electrons in the accretion disk of the AGN NGC 4261 \cite{Jones_01}
which leads to $T_p \approx 10^{-6}s$. 
The low average magnetic field in the accretion disk gives rise to a
weak ESA capable of producing relativistic electrons from the thermal
background over an acceleration time of a few thousand $T_p$, i.e. a few
milliseconds. The rising time of synchrotron emissions would then be of
the same order.
Such a mechanism 
may explain the rapid fluctuations in the synchrotron emission from 
accretion disks of microquasars \cite{Fender_04} or from GRBs \cite{Piran_04}. 
However, we have to point out that our plasma initial conditions are 
idealized and, in addition, only limited data is currently available 
about typical plasma parameters of microquasars accretion disks or gamma ray bursts.  
The present work constitutes a first step
towards a kinetic model of electron acceleration in the foreshock
region of mildly relativistic astrophysical shocks. It shows that
ESA is a promising candidate for the rapid generation of highly 
relativistic electrons in astrophysical environments that support
such shocks.
Future work should consider ESA driven by ion beams that have evolved
self-consistently out of a shock, together with the effects of oblique magnetic
field angles.

\ack
This work was supported in part by: the European Commission through
the Grant No. HPRN-CT-2001-00314; the Deutsche Forschungsgemeinschaft (DFG);
 the Engineering and Physical
Sciences Research Council (EPSRC): and the United Kingdom Atomic
Energy Authority (UKAEA). The authors thank the Swedish National
Supercomputer Centre (NSC), the Center for Parallel Computers (PDC)
at Stockholm and the Centre for Scientific Computing (CSC) at the University of Warwick, with support from Science Research Investment Fund grant (grant code TBA), for the provision of  computer time.
N J Sircombe would like to thank Padma Shukla and the rest of the Institut f\"ur Theoretische Physik IV at the Ruhr-Universit\"at Bochum for their kind hospitality during his stay.

\newpage
\bibliographystyle{unsrt}

\begin{thebibliography}{10}

\bibitem{Bingham_04a} Bingham R, Mendonca J T and Shukla P K 2004
{\it Plasma Phys. Contr. Fusion} {\bf 46} R1

\bibitem{Bingham_04b} Bingham R, Kellet B J, Bryans P, Summers H P,
Torney M, Shapiro V D, Spicer D S and O'Brien M 2004 
{\it Astrophys. J.} {\bf 601} 896

\bibitem{Sircombe_05} Sircombe N J, Arber T D and Dendy R O 2005 
{\it Phys. Plasmas} {\bf 12} 012303

\bibitem{Bingham_03} Bingham R 2003 {\it Nature} {\bf 424} 258

\bibitem{Dieckmann_00} Dieckmann M E, Ljung P, Ynnerman A and McClements
K G 2000 {\it Phys. Plasmas} {\bf 7} 5171

\bibitem{Kuramitsu_05} Kuramitsu Y and Krasnoselskikh 2005
{\it Phys. Rev. Lett.} {\bf 94} 031102

\bibitem{Sagdeev_73} Sagdeev R Z and Shapiro V D 1973 {\it JETP Lett.} {\bf 17} 279 

\bibitem{Katsouleas_83} Katsouleas T and Dawson J M 1983 {\it Phys. Rev. Lett.} {\bf 51} 392
\bibitem{Luque_05} Luque A and Schamel H 2005 {\it Phys. Rep.} {\bf 415} 261

\bibitem{Dieckmann_04a} Dieckmann M E, Eliasson B and Shukla P K 2004 
{\it Astrophys. J} {\bf 617} 1361

\bibitem{Fender_04} Fender R and Belloni T 2004 {\it Annu. Rev. Astron. 
Astrophys.} {\bf 42} 317

\bibitem{Piran_04} Piran T 2004 {\it Rev. Mod. Phys.} {\bf 76} 1143

\bibitem{Buneman_58} Buneman O 1958 {\it Phys. Rev. Lett.} {\bf 1} 8

\bibitem{Thode_73} Thode L E and Sudan R N 1973 {\it Phys. Rev. Lett.}
{\bf 30} 732

\bibitem{Bauer92} Bauer F and Schamel H 1992 {\it Physica D} {\bf 54} 235

\bibitem{Galeev_84} Galeev A A 1984 {\it Sov. Phys. J. Exp. Theor. Phys.}
{\bf 59} 965

\bibitem{McClements_97} McClements K G, Dendy R O, Bingham R, Kirk J G and 
Drury L O 1997 {\it Mon. Not. R. Astron. Soc.} {\bf 291} 241

\bibitem{Shapiro_03} Shapiro V D and Ucer D 2003 {\it Planet. Space Sci.} {\bf 51} 665

\bibitem{Ucer_01} Ucer D and Shapiro V D 2001 {\it Phys. Rev. Lett.} {\bf 87} 075001

\bibitem{Lee_96} Lee M, A Shapiro V D and Sagdeev R Z 1996 
         {\it J. Geophys. Res.} {\bf 101} 4777

\bibitem{Lee_04} Lee R, Chapman S C and Dendy R O 2004 {\it Astrophys J.}
	 {\bf 604} 187

\bibitem{Lee_05} Lee R, Chapman S C and Dendy R O 2004 {\it Phys. Plasmas}
	 {\bf 12} 012901

\bibitem{Schmitz_02a} Schmitz H, Chapman S C and Dendy R O 2002a {\it Astrophys. J.}
        {\bf 570} 637

\bibitem{Schmitz_02b} Schmitz H, Chapman S C and Dendy R O 2002b {\it Astrophys. J.}
        {\bf 579} 327
   
\bibitem{Shimada_00} Shimada N and Hoshino M 2000 {\it Astrophys. J.} {\bf 543} L67

\bibitem{Shimada_05} Shimada N and Hoshino M 2005 {\it J. Geophys. Res.} {\bf 110}
        A02105

\bibitem{Hoshino_02} Hoshino M and Shimada N 2002 {\it Astrophys. J.} {\bf 572} 880

\bibitem{Aoki_04} Aoki S I, Koide S, Kudoh T, Nakayama K and Shibata K 2004
        {\it Astrophys. J.} {\bf 610} 897


\bibitem{Dieckmann_04b} Dieckmann M E, Eliasson B and Shukla P K 2004 {\it Phys. Plasmas}
        {\bf 11} 1394

\bibitem{Dieckmann_04c} Dieckmann M E, Eliasson B, Stathopoulos A and Ynnerman A 2004
        {\it Phys. Rev. Lett.} {\bf 92} 065006

\bibitem{Schamel_96} Schamel H and Korn J 1996 {\it Phys. Scripta} {\bf T63} 63

\bibitem{McClements_01} McClements K G, Dieckmann M E, Ynnerman A, Chapman 
S C and Dendy R O 2001 {\it Phys. Rev. Lett.} {\bf 87} 255002

\bibitem{Gohda_04} Gohda T, Ishiguro S, Iizuka S and Sato N {\it Phys. Rev. Lett.} {\bf 92} 045002

\bibitem{Sack_1987} Sack Ch and Schamel H 1987 {\it Phys. Rep.} {\bf 156} 311

\bibitem{Eastwood_91} Eastwood J W 1991 {\it Comput. Phys. Comm.} {\bf 92} 252

\bibitem{Arber_02} Arber T D and Vann R G L 2002 {\it J. Comp. Phys.} {\bf 180} 339

\bibitem{Lembege_04} Lembege B et al. 2004 {\it Space Sci. Rev.} {\bf 110} 161

\bibitem{Jones_01} Jones D L, Wehrle A E, Piner B G and Meier D L 2001 
        {\it Astrophys. J.} {\bf 553} 968

\bibitem{Dieckmann_04d} Dieckmann M E, Ynnerman A, Chapman S C, Rowlands G
and Andersson N 2004 {\it Phys. Scripta} {\bf 69} 456

\bibitem{Escande_82} Escande D F 1982 {\it Phys. Scripta} {\bf T2} 126

\bibitem{Schamel_86} Schamel H 1986 {\it Phys. Rep.} {\bf 140} 161

\bibitem{Eliasson_05} Eliasson B and Shukla P K 2005 {\it Nonlinear Proc. Geophys.} {\bf 12} 269

\bibitem{Bernstein_57} Bernstein I B, Greene, J M \&
      Kruskal M D 1957 {\it Phys. Rev. Lett.} {\bf 108} 546

\bibitem{Kruer_69} Kruer W L, Dawson J M and Sudan R N 1969 {\it Phys. Rev. Lett.}
        {\bf 23} 838
\bibitem{Krasovsky_94} Krasovsky V L 1994 {\it Phys. Scripta} {\bf 49} 489

\bibitem{Schwarzmeier_79} Schwarzmeier J L, Lewis H R, Abraham-Shrauner B 
and Simon K R 1979 {\it Phys. Fluids} {\bf 22} 1747 

\bibitem{Korn_96}  Korn J and Schamel H
      1996, {\it J. Plasma Phys.} {\bf 56} 339

\bibitem{Morales_74} Morales G J and Lee Y C 1974 {\it Phys. Rev. Lett.} 
{\bf 33} 1534

\bibitem{Dieckmann_99} Dieckmann M E, Chapman S C, Dendy R O and Drury L O C
2000 {\it Astron. Astrophys} {\bf 356} 377

\bibitem{Karney_77} Karney C F F and Bers A 1977 {\it Phys. Rev. Lett.} {\bf 39} 550
      
\end{thebibliography}

\end{document}